\documentclass[review]{elsarticle}
\usepackage{color}
\usepackage{hyperref}
\journal{Journal of \LaTeX\ Templates}
\usepackage{natbib}

\begin{document}

\begin{frontmatter}

\title{Lambda beta-decay in-medium}

\author{P.~A.~M.~Guichon}
\address{DPhN-IRFU, CEA Saclay, F91191 Gif sur Yvette, France}
\author{A.~W.~Thomas}
\address{CSSM and ARC Centre of Excellence for Particle Physics at the Terascale, Department of Physics, University of Adelaide SA 5005 Australia}




\begin{abstract}
Under the working hypothesis that the structure of a bound hadron is modified by its interactions with other hadrons, one may expect to see  changes in carefully chosen observables. In the light of a recent proposal to measure the axial charge in the strangeness changing beta-decay of a bound Lambda hyperon, we examine the size of the change expected within the quark-meson coupling model. It is predicted to be significant.  
\end{abstract}


\end{frontmatter}

%\linenumbers

%%%%%%%%%%%
\section{Introduction}
%%%%%%%%%%%

After more than a hundred years the search for a fundamental theory of nuclear structure is continuing. The traditional approach, based upon many-body theory using phenomenological nucleon-nucleon (NN) forces with parameters determined by fits to NN data and supplemented by a three-body force with parameters constrained by nuclear data, has proven very successful for light nuclei~\cite{Pieper:2001mp}. Chiral effective field theory, which builds upon the symmetries of QCD~\cite{Weinberg:1991um}, is currently widely used and also successful for lighter nuclei~\cite{Epelbaum:2000mx}. For heavy nuclei the density functional approach, pioneered by Brink and Vautherin~\cite{Vautherin:1971aw}, is extremely widely used with hundreds of Skyrme forces in the literature~\cite{Dutra:2012mb}. Until recently these Skyrme forces have been purely phenomenological, with a sizeable number of parameters determined by fitting selected nuclear data. A common thread in all of these approaches is that the particles which feel these phenomenological forces have the same structure as free protons and neutrons.

Of course, at the present time no-one doubts that the correct underlying theory of nuclear structure is QCD and the fundamental degrees of freedom are quarks and gluons. In spite of the substantial advances being made in lattice QCD, there is as yet no way to calculate the properties of a nucleus like Uranium directly from QCD. Chiral effective field theory at least guarantees the symmetries of QCD but is currently limited to neutrons, protons and pions as the relevant degrees of freedom. We will be concerned with an alternate approach, which starts with confined quarks as the degrees of freedom. This model, known as the quark-meson coupling (QMC)                                           model~\cite{Guichon:1987jp,Guichon:1995ue,Saito:2005rv} has been developed over some decades, with successful applications to a variety of phenomena; from traditional nuclear structure, to hypernuclei~\cite{Guichon:2008zz}, $\omega$~\cite{Tsushima:1998qw},         $\eta$ and $\eta^\prime$~\cite{Bass:2005hn,Bass:2010kr,Nanova:2016cyn,Metag:2017yuh} mesons bound in matter and the EMC effect~\cite{Thomas:1989vt,Cloet:2006bq}, to cite just a few examples. Like the other approaches the QMC model reduces the many-quark problem to a problem involving many clusters of confined quarks. However, within the QMC model the structure of those clusters is self-consistently determined by the interactions of the confined quarks in them with the quarks in the neighbouring clusters. 

In this approach the relativistic character of the mean fields is critical, with a scalar field producing very different effects on the internal dynamics of the clusters from those generated by a Lorentz vector mean field. Indeed, while the latter primarily redefines the energy and contributes to the spin-orbit force, the former modifies the Dirac wave functions of the valence quarks, enhancing their lower components and as a consequence opposing the applied scalar field; an effect parametrized in terms of a ''scalar polarizability''. In the QMC model the scalar polarizability is the prime mechanism leading to the saturation of nuclear matter. The connection between the modification of the internal structure of the clusters of quarks with nucleon quantum numbers and either three-body forces or density dependent effective forces was established a decade ago by constructing a density functional equivalent to the underlying quark theory~\cite{Guichon:2004xg}. However, it is only in the past year or so that the density dependent force derived from the QMC model~\cite{Guichon:2006er}, which itself has just a few parameters, was shown to produce a remarkably accurate description of the properties of atomic nuclei across the periodic table~\cite{Stone:2016qmi}.

The recent developments of the QMC model have established that it is capable of providing a realistic description of nuclear properties. Yet underlying this more or less conventional density functional (the detailed form of the non-linearity of the derived density dependence is rather less conventional) is the prediction that the structure of the bound ''nucleon'' is changed by the interactions with the local medium. Within an extension of the approach based upon the NJL model~\cite{Bentz:2001vc} rather than the MIT bag, this has been shown to lead to a rather satisfactory description of the EMC effect, with predictions for an enhanced EMC effect in the spin structure   function~\cite{Cloet:2006bq} and observable consequences for parity violating deep inelastic scattering~\cite{Cloet:2012td}. It also leads to a dramatic correction to the Coulomb sum rule~\cite{Cloet:2015tha}. All of these predictions will be subject to experimental tests but given the dramatic revision of our view of nuclear structure implied by the model it is vital to test the predictions in as many different ways as possible. 

Here we investigate a new signal of the predicted change in the structure of a bound nucleon which will hopefully be the subject of experimental investigation at J-PARC. In particular, we consider the change in the $\Delta S = 1$ axial charge which is involved in the beta-decay of a bound $\Lambda$-hyperon. Some decay modes of the $\Lambda$ hyperon in nuclei have already been studied, of course. The mesonic and non-mesonic decay rates in $^5$He, $^{12}$C and various other nuclei have been studied at BNL, KEK and DAFNE and theoretical calculations for those decay rates have been performed within a meson exchange picture~\cite{Botta:2015lms,Itonaga:2010zz,Parreno:2007zz}. However, there has so far been neither a calculation nor a measurement of the beta-decay of $\Lambda$ hypernuclei. This work is inspired in part by discussions with H. Tamura who is planning to propose such an experiment at 
J-PARC~\cite{Tamura}. Of course, there is a great deal of interest already in the modification of the axial charge of the nucleon in-medium, especially as this quantity enters to the fourth power in the lifetimes associated with neutrinoless double beta-decay, which provides a promising window onto whether or not the neutrino is a Majorana particle~\cite{Vergados:2016hso}. By focussing on the beta-decay of a bound $\Lambda$-hyperon in nuclei with different mass numbers, one has the opportunity to separate the intrinsic change in the corresponding axial charge from other potential corrections, such as meson exchange currents. 

%%%%%%%%%%%%%%%%%%%%%%%%%%%%%
\section{Modification of the $\Delta S=1$ axial charge}
%%%%%%%%%%%%%%%%%%%%%%%%%%%%%

The calculation of the mean scalar and vector potentials within the QMC model has been explained in a number of places. Motivated by the Zweig rule, the model includes {\em only} the coupling of the $\sigma , \omega$ and 
$\rho$ mesons to the $u$ and $d$ quarks, not the strange quark. This has proven remarkably successful in describing the properties of hypernuclei. We fix the coupling constants of these mesons to the light quarks so as to reproduce the saturation properties of symmetric nuclear matter, as well as the symmetry energy at saturation. 

For the light quark in the final nucleon (following beta-decay) the quark has an effective mass modified by its coupling to the mean scalar field. This effective mass, $m^{*}$, is negative at all but the lowest densities but this creates no problems as the quark is confined in a cavity of radius $R$ and the eigenenergy of the quark is positive.
Denoting this eigenenergy as $\Omega_{\alpha}/R$, the corresponding Dirac wave function is
\begin{equation}
\phi_{\alpha}=\left(\begin{array}{c}
f_{\alpha}(r)\\
i\vec{\sigma}.\hat{r}g_{\alpha}(r)\end{array}\right)\frac{\chi}{\sqrt{4\pi}}
\label{eq:phi}
\end{equation}
with
\begin{eqnarray*}
&&f_{\alpha}(r)  ={\cal N}_{\alpha} \frac{1}{r}\sin\left[\sqrt{\Omega_{\alpha}^{2}-(m^{*}R)^{2}}\frac{r}{R}\right]\\
&&g_{\alpha}(r)  ={\cal N}_{\alpha}
\frac{R}{(\Omega_{\alpha}+m^{*}R)r}\\
&&\left(\frac{1}{r}
\sin\left[\sqrt{\Omega_{\alpha}^{2}-(m^{*}R)^{2}}
\frac{r}{R}\right]-\frac{\sqrt{\Omega_{\alpha}^{2}-(m^{*}R)^{2}}}{R}
\cos\left[\sqrt{\Omega_{\alpha}^{2}-(m^{*}R)^{2}}
\frac{r}{R}\right]\right) \, .
\end{eqnarray*}
Here ${\cal N}$ is the normalization constant such that
\[
\int_{0}^{R}r^{2}dr\left(f_{\alpha}^{2}+g_{\alpha}^{2}\right)=1 \, .
\]
The eigenvalue is determined by the boundary condition $f(R)=g(R)$.

Because there is no coupling of the mean scalar field to the strange quark in the $\Lambda$ hyperon, its wave function is exactly as given in the original MIT bag model. This wave function, which we write as
\begin{equation}
\psi_{\alpha}=\left(\begin{array}{c}
\tilde{f}_{\alpha}(r)\\
i\vec{\sigma}.\hat{r}\tilde{g}_{\alpha}(r)\end{array}\right)\frac{\chi}{\sqrt{4\pi}}
\label{eq:psi}
\end{equation}
has exactly the same form as $\phi$ above but with the strange quark mass, $m_s$, replacing $m^{*}$ and an appropriate normalization constant,  
$\tilde{{\cal N}}$.

Following a trivial calculation, the strangeness changing axial charge is then given by:
\begin{equation}
g_A(\Delta S = 1) = \int_{0}^{R}r^{2}dr\left( \tilde{f}(r) f(r) - \frac{1}{3} \tilde{g}(r) g(r)\right) \, .
\label{eq:ga}
\end{equation}
To estimate the change in this axial charge arising from the change in the internal structure of the bound nuclon we recall that the quark effective mass in-medium depends on the  $\sigma$ field according to $m^{*} = m - g_{\sigma q} \sigma$. Here $m$ is the light quark mass appearing in the MIT bag model Lagrangian density and, as it is only a few MeV, we set it to zero.  The quark coupling is related to the (free) nucleon coupling, $g_{\sigma N}$, through
\begin{equation}
g_{\sigma N}=3 g_{\sigma q}\int_{0}^{R} d\vec r \bar{\phi}\phi=3\times 0.479 \, g_{\sigma q} \, .
\label{eq:qqqq}
\end{equation}

For the $\sigma$ field we use a local density approximation which gives~\cite{Guichon:2006er}
\begin{equation}
g_{\sigma N}\sigma = \frac{G_{\sigma} \rho}{(1+d G_{\sigma} \rho)} \, , 
\label{eq:pppp}
\end{equation}
with $\rho$ the local density and $G_{\sigma}=g_{\sigma N}^{2}/m_\sigma^2$. In the above expression we have neglected small relativistic effects and retained the dominant  many-body effect arising from the scalar polarisability, $d$~\cite{Guichon:2006er}. The integrals in Eq.~(\ref{eq:ga})   are computed assuming that the bag radius is the same for the light and strange quark. This is a standard approximation which is sufficient for our purpose. The coupling $G_{\sigma}$ is chosen so as to reproduce the saturation properties of nuclear matter. Since this depends on the selected approximations (Hartree vs Hartree Fock, relativistic vs non relativistic and so on), we show results for  three values of $G_\sigma$ which cover the range found over the various approximations. The bag radius is set to 1fm, from which we calculate $d=0.18{\rm fm}$.  

\begin{figure}
\centering
    \includegraphics[scale=0.5]{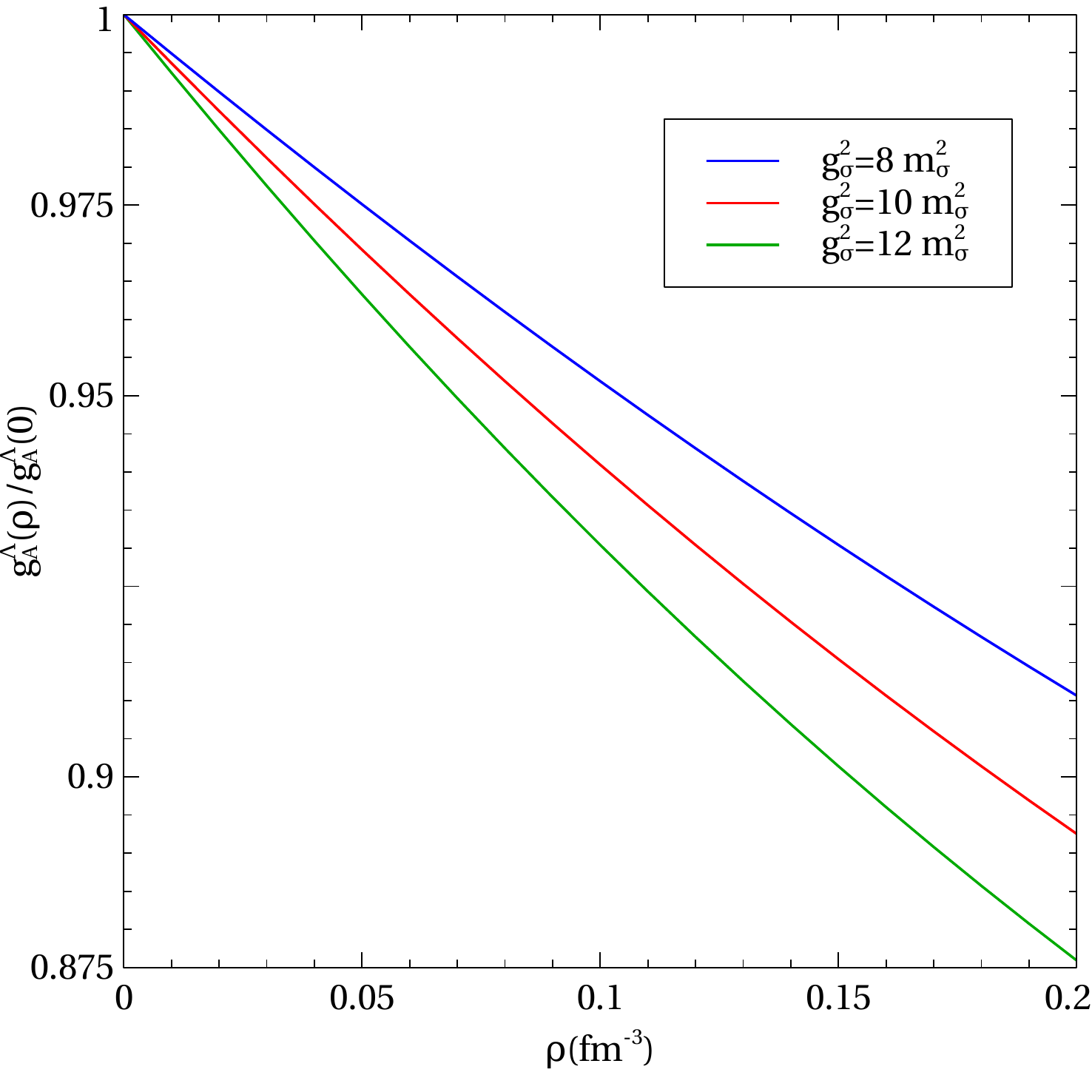}
    \caption{Relative variation of the axial coupling as a function of density}
    \label{fig: gaLambda}
  \end{figure}

The relative change in the axial charge  is illustrated in Fig. \ref{fig: gaLambda}  as a function of density. The suppression is $9 \pm 1.6$ \% at nuclear matter density. This corresponds to a reduction of up to 20\% in the beta-decay rate for the bound $\Lambda$-hyperon.
%

%%%%%%%%%%%%%%%%%%%%%%%%%%%%%%
\section{Modification of the $\Delta S=1$ vector charge}
%%%%%%%%%%%%%%%%%%%%%%%%%%%%%%

Our primary focus is on the change of the axial charge because of the protection accorded to the vector charge by the Ademollo-Gatto theorem~\cite{Ademollo:1964sr}. For example, in the case of neutron beta-decay the deviation of the vector charge from unity goes as the square of the small $u$-$d$ mass difference, $(m_u-m_d)^2$. For the $\Lambda$-hyperon it is proportional to $(m_s-m_u)^2$ in free space and $(m_s - m_u +g_\sigma^q \sigma)^2$ in-medium. It is straightforward to generalise the calculation for neutron beta-decay (in free space and in-medium) presented in Ref.~\cite{Guichon:2011gc} to the present case, with the leading correction to the vector charge given by
\begin{equation}
\delta g_V = - \frac{(m_s + g_\sigma^q \sigma)^2}{2} \sum_{\alpha \neq 0} \frac{I_\alpha^2}{(M_0-M_\alpha)^2} \, .
\label{eq:vec}
\end{equation}
Here the integral $I_\alpha$ is
\begin{equation}
\int_{bag} \phi_\alpha^\dagger \gamma^0 \phi_0 \, ,
\label{eq:Ialpha}
\end{equation}
and $\phi_\alpha$ are the bag wave functions at zero quark mass in free space ($\alpha=0$  is the ground-state). The result for the relative variation of the vector coupling  is illustrated in Fig.~2, as a function of density. The variation from free space ($\rho=0$) to nuclear matter density is around 3.5\%. This is not insignificant but much smaller than the reduction in the axial charge, which also carries a weighting approximately a factor of three larger~\cite{Garcia:1985xz}.
\begin{figure}
\centering
    \includegraphics[scale=0.5]{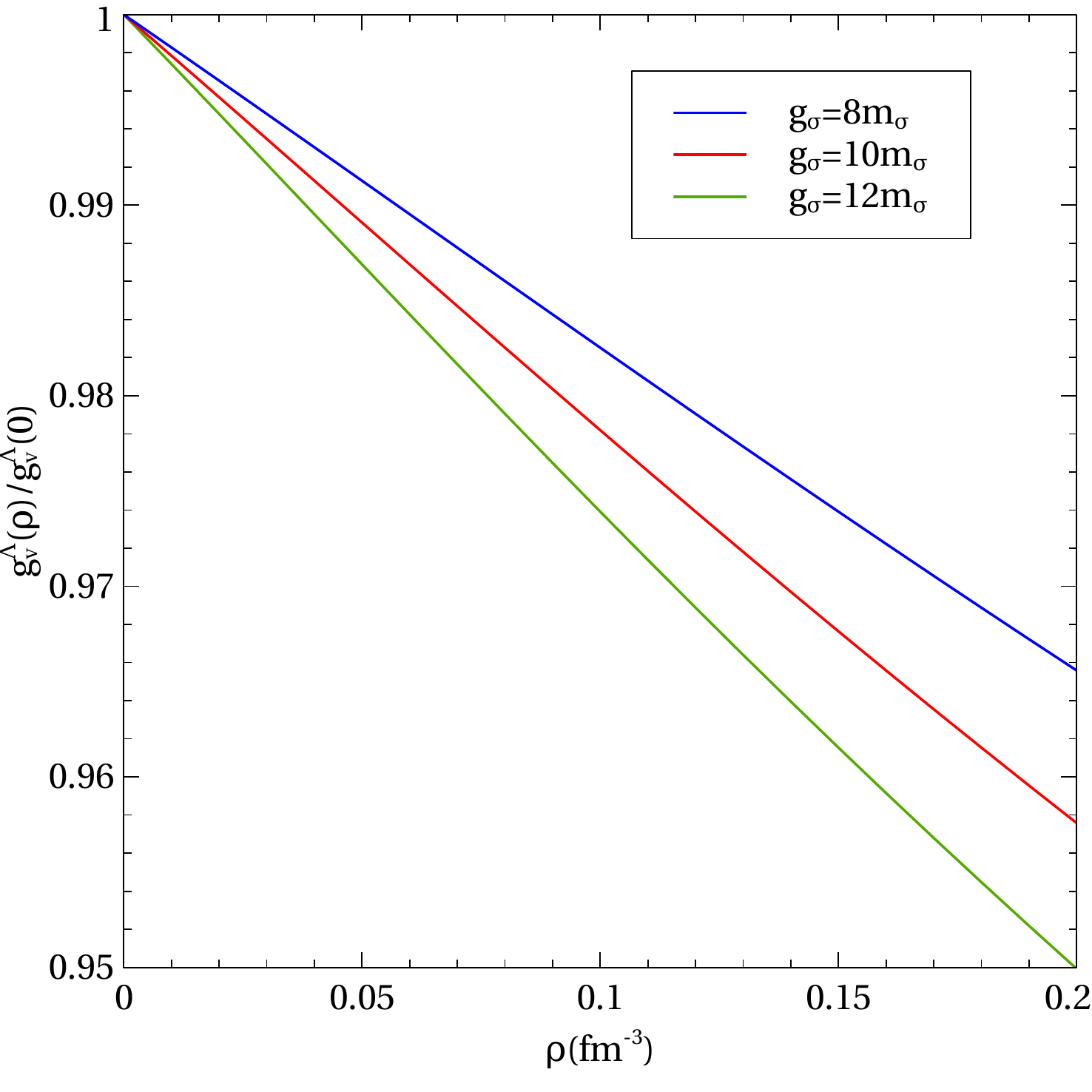}
    \caption{Relative variation of the vector  coupling as a function of density for $m_s=300MeV$}
    \label{fig: gvLambda}
  \end{figure}
%

%%%%%%%%%%%%%%%%
\section{Concluding remarks}
%%%%%%%%%%%%%%%%

If the structure of a bound nucleon is modified, as suggested in the quark-meson coupling model, there will be a sizeable correction to the strangeness changing axial charge of the $\Lambda$ hyperon, which should be measureable in the beta-decay of $\Lambda$-hypernuclei. Indeed, at nuclear matter density the suppression  is a as large as 10\%, which would correspond to a suppression of order 20\%   of the partial decay width. Even at the lower densities appropriate to the nuclear surface the change in the axial charge is predicted to be as much as 5\%. There is a smaller reduction (of order 3.5\%) in the corresponding vector charge. Establishing the existence of this reduction would provide strong supporting evidence for the proposition that the structure of a bound nucleon is altered significantly in-medium, with all the associated consequences for our understanding of nuclear structure.

A complete analysis of the proposed experiment also requires an accurate calculation of the Pauli suppression of $\Lambda$-hyperon beta-decay. At an average density between 50 and 100\% of $\rho_0$, the Pauli suppression of the decay rate, compared with free space, should be around a factor of five, so it must be calculated accurately. On the other hand, this is a straightforard calculation and should be very reliable.

One of the beauties of hypernuclei is that one can choose to look at data sets where the decaying hyperon is either located near the centre of the nucleus, for example in a 1s-state, or in a higher lying state, with correspondingly lower density. This should be a powerful tool for testing this proposal, which goes to the heart of our understanding of nuclear structure.

\section*{Acknowledgements}
We would like to acknowledge helpful discussions with H. Tamura. This work was supported by the University of Adelaide and by the Australian Research Council through the ARC Centre of Excellence for Particle Physics at the Terascale (CE110001104) and Discovery Project DP150103164.

\section*{References}

%\bibliography{mybibfile}

\begin{thebibliography}{55}
%
%
%\cite{Pieper:2001mp}
\bibitem{Pieper:2001mp}
  S.~C.~Pieper and R.~B.~Wiringa,
  %``Quantum Monte Carlo calculations of light nuclei,''
  Ann.\ Rev.\ Nucl.\ Part.\ Sci.\  {\bf 51} (2001) 53
  doi:10.1146/annurev.nucl.51.101701.132506
  [nucl-th/0103005].
  %%CITATION = doi:10.1146/annurev.nucl.51.101701.132506;%%
%
%\cite{Weinberg:1991um}
\bibitem{Weinberg:1991um}
  S.~Weinberg,
  %``Effective chiral Lagrangians for nucleon - pion interactions and nuclear forces,''
  Nucl.\ Phys.\ B {\bf 363} (1991) 3.
  doi:10.1016/0550-3213(91)90231-L
  %%CITATION = doi:10.1016/0550-3213(91)90231-L;%%
%
%\cite{Epelbaum:2000mx}
\bibitem{Epelbaum:2000mx}
  E.~Epelbaum, H.~Kamada, A.~Nogga, H.~Witala, W.~Gloeckle and U.~G.~Meissner,
  %``The Three nucleon and four nucleon systems from chiral effective field theory,''
  Phys.\ Rev.\ Lett.\  {\bf 86} (2001) 4787
  doi:10.1103/PhysRevLett.86.4787
  [nucl-th/0007057].
  %%CITATION = doi:10.1103/PhysRevLett.86.4787;%%
%
%\cite{Vautherin:1971aw}
\bibitem{Vautherin:1971aw}
  D.~Vautherin and D.~M.~Brink,
  %``Hartree-Fock calculations with Skyrme's interaction. 1. Spherical nuclei,''
  Phys.\ Rev.\ C {\bf 5} (1972) 626.
  doi:10.1103/PhysRevC.5.626
  %%CITATION = doi:10.1103/PhysRevC.5.626;%%
%
%\cite{Dutra:2012mb}
\bibitem{Dutra:2012mb}
  M.~Dutra, O.~Lourenco, J.~S.~Sa Martins, A.~Delfino, J.~R.~Stone and P.~D.~Stevenson,
  %``Skyrme Interaction and Nuclear Matter Constraints,''
  Phys.\ Rev.\ C {\bf 85} (2012) 035201
  doi:10.1103/PhysRevC.85.035201
  [arXiv:1202.3902 [nucl-th]].
  %%CITATION = doi:10.1103/PhysRevC.85.035201;%%
%
%\cite{Guichon:1987jp}
\bibitem{Guichon:1987jp}
  P.~A.~M.~Guichon,
  %``A Possible Quark Mechanism for the Saturation of Nuclear Matter,''
  Phys.\ Lett.\ B {\bf 200} (1988) 235.
  doi:10.1016/0370-2693(88)90762-9
  %%CITATION = doi:10.1016/0370-2693(88)90762-9;%%
%
%\cite{Guichon:1995ue}
\bibitem{Guichon:1995ue}
  P.~A.~M.~Guichon, K.~Saito, E.~N.~Rodionov and A.~W.~Thomas,
  %``The Role of nucleon structure in finite nuclei,''
  Nucl.\ Phys.\ A {\bf 601} (1996) 349
  doi:10.1016/0375-9474(96)00033-4
  [nucl-th/9509034].
  %%CITATION = doi:10.1016/0375-9474(96)00033-4;%%
%
%\cite{Saito:2005rv}
\bibitem{Saito:2005rv}
  K.~Saito, K.~Tsushima and A.~W.~Thomas,
  %``Nucleon and hadron structure changes in the nuclear medium and impact on observables,''
  Prog.\ Part.\ Nucl.\ Phys.\  {\bf 58} (2007) 1
  doi:10.1016/j.ppnp.2005.07.003
  [hep-ph/0506314].
  %%CITATION = doi:10.1016/j.ppnp.2005.07.003;%%
%
%\cite{Guichon:2008zz}
\bibitem{Guichon:2008zz}
  P.~A.~M.~Guichon, A.~W.~Thomas and K.~Tsushima,
  %``Binding of hypernuclei in the latest quark-meson coupling model,''
  Nucl.\ Phys.\ A {\bf 814} (2008) 66
  doi:10.1016/j.nuclphysa.2008.10.001
  [arXiv:0712.1925 [nucl-th]].
  %%CITATION = doi:10.1016/j.nuclphysa.2008.10.001;%%
%
%\cite{Tsushima:1998qw}
\bibitem{Tsushima:1998qw}
  K.~Tsushima, D.~H.~Lu, A.~W.~Thomas and K.~Saito,
  %``Are eta and omega nuclear states bound?,''
  Phys.\ Lett.\ B {\bf 443} (1998) 26
  doi:10.1016/S0370-2693(98)01336-7
  [nucl-th/9806043].
  %%CITATION = doi:10.1016/S0370-2693(98)01336-7;%%
%
%\cite{Bass:2005hn}
\bibitem{Bass:2005hn}
  S.~D.~Bass and A.~W.~Thomas,
  %``eta bound states in nuclei: A Probe of flavor-singlet dynamics,''
  Phys.\ Lett.\ B {\bf 634} (2006) 368
  doi:10.1016/j.physletb.2006.01.071
  [hep-ph/0507024].
  %%CITATION = doi:10.1016/j.physletb.2006.01.071;%%
%
%\cite{Bass:2010kr}
\bibitem{Bass:2010kr}
  S.~D.~Bass and A.~W.~Thomas,
  %``$\eta - \eta'$ mixing in $\eta$-mesic nuclei,''
  Acta Phys.\ Polon.\ B {\bf 41} (2010) 2239
  [arXiv:1007.0629 [hep-ph]].
  %%CITATION = ARXIV:1007.0629;%%
%
%\cite{Nanova:2016cyn}
\bibitem{Nanova:2016cyn}
  M.~Nanova {\it et al.} [CBELSA/TAPS Collaboration],
  %``Determination of the real part of the $\eta$'-Nb optical potential,''
  Phys.\ Rev.\ C {\bf 94} (2016) no.2,  025205
  doi:10.1103/PhysRevC.94.025205
  [arXiv:1607.07228 [nucl-ex]].
  %%CITATION = doi:10.1103/PhysRevC.94.025205;%%
%
%\cite{Metag:2017yuh}
\bibitem{Metag:2017yuh}
  V.~Metag, M.~Nanova and E.~Y.~Paryev,
  %``Meson-nucleus potentials and the search for meson-nucleus bound states,''
  arXiv:1706.09654 [nucl-ex].
  %%CITATION = ARXIV:1706.09654;%%
%
%\cite{Thomas:1989vt}
\bibitem{Thomas:1989vt}
  A.~W.~Thomas, A.~Michels, A.~W.~Schreiber and P.~A.~M.~Guichon,
  %``A New Approach To Nuclear Structure Functions,''
  Phys.\ Lett.\ B {\bf 233} (1989) 43.
  doi:10.1016/0370-2693(89)90612-6
  %%CITATION = doi:10.1016/0370-2693(89)90612-6;%%
%
%\cite{Bentz:2001vc}
\bibitem{Bentz:2001vc}
  W.~Bentz and A.~W.~Thomas,
  %``The Stability of nuclear matter in the Nambu-Jona-Lasinio model,''
  Nucl.\ Phys.\ A {\bf 696} (2001) 138
  doi:10.1016/S0375-9474(01)01119-8
  %[nucl-th/0105022].
  %%CITATION = doi:10.1016/S0375-9474(01)01119-8;%%
%
%\cite{Cloet:2006bq}
\bibitem{Cloet:2006bq}
  I.~C.~Clo\"et, W.~Bentz and A.~W.~Thomas,
  %``EMC and polarized EMC effects in nuclei,''
  Phys.\ Lett.\ B {\bf 642} (2006) 210
  doi:10.1016/j.physletb.2006.08.076
  [nucl-th/0605061].
  %%CITATION = doi:10.1016/j.physletb.2006.08.076;%%
%
%\cite{Guichon:2004xg}
\bibitem{Guichon:2004xg}
  P.~A.~M.~Guichon and A.~W.~Thomas,
  %``Quark structure and nuclear effective forces,''
  Phys.\ Rev.\ Lett.\  {\bf 93} (2004) 132502
  doi:10.1103/PhysRevLett.93.132502
  [nucl-th/0402064].
  %%CITATION = doi:10.1103/PhysRevLett.93.132502;%%
%
%\cite{Guichon:2006er}
\bibitem{Guichon:2006er}
  P.~A.~M.~Guichon, H.~H.~Matevosyan, N.~Sandulescu and A.~W.~Thomas,
  %``Physical origin of density dependent force of the skyrme type within the quark meson coupling model,''
  Nucl.\ Phys.\ A {\bf 772} (2006) 1
  doi:10.1016/j.nuclphysa.2006.04.002
  [nucl-th/0603044].
  %%CITATION = doi:10.1016/j.nuclphysa.2006.04.002;%%
%
%\cite{Stone:2016qmi}
\bibitem{Stone:2016qmi}
  J.~R.~Stone, P.~A.~M.~Guichon, P.~G.~Reinhard and A.~W.~Thomas,
  %``Finite Nuclei in the Quark-Meson Coupling Model,''
  Phys.\ Rev.\ Lett.\  {\bf 116} (2016) no.9,  092501
  doi:10.1103/PhysRevLett.116.092501
  [arXiv:1601.08131 [nucl-th]].
  %%CITATION = doi:10.1103/PhysRevLett.116.092501;%%
%
%\cite{Cloet:2012td}
\bibitem{Cloet:2012td}
  I.~C.~Clo\"et, W.~Bentz and A.~W.~Thomas,
  %``Parity-violating DIS and the flavour dependence of the EMC effect,''
  Phys.\ Rev.\ Lett.\  {\bf 109} (2012) 182301
  doi:10.1103/PhysRevLett.109.182301
  [arXiv:1202.6401 [nucl-th]].
  %%CITATION = doi:10.1103/PhysRevLett.109.182301;%%
%
%\cite{Cloet:2015tha}
\bibitem{Cloet:2015tha}
  I.~C.~Clo\"et, W.~Bentz and A.~W.~Thomas,
  %``Relativistic and Nuclear Medium Effects on the Coulomb Sum Rule,''
  Phys.\ Rev.\ Lett.\  {\bf 116} (2016) no.3,  032701
  doi:10.1103/PhysRevLett.116.032701
  [arXiv:1506.05875 [nucl-th]].
  %%CITATION = doi:10.1103/PhysRevLett.116.032701;%%
%
%\cite{Botta:2015lms}
\bibitem{Botta:2015lms}
  E.~Botta, T.~Bressani, S.~Bufalino and A.~Feliciello,
  %``Status and perspectives of experimental studies on hypernuclear weak decays,''
  Riv.\ Nuovo Cim.\  {\bf 38} (2015) no.9,  387.
  doi:10.1393/ncr/i2015-10116-x
  %%CITATION = doi:10.1393/ncr/i2015-10116-x;%%
%
%\cite{Itonaga:2010zz}
\bibitem{Itonaga:2010zz}
  K.~Itonaga and T.~Motoba,
  %``Hypernuclear weak decays,''
  Prog.\ Theor.\ Phys.\ Suppl.\  {\bf 185} (2010) 252.
  doi:10.1143/PTPS.185.252
  %%CITATION = doi:10.1143/PTPS.185.252;%%
%
%\cite{Parreno:2007zz}
\bibitem{Parreno:2007zz}
  A.~Parreno,
  %``Weak decays of hypernuclei,''
  Lect.\ Notes Phys.\  {\bf 724} (2007) 141.
 % doi:10.1007/978-3-540-72039-3_5
  %%CITATION = doi:10.1007/978-3-540-72039-3_5;%%
%
\bibitem{Tamura}
H. Tamura, private communication.
%
%\cite{Vergados:2016hso}
\bibitem{Vergados:2016hso}
  J.~D.~Vergados, H.~Ejiri and F.~Šimkovic,
  %``Neutrinoless double beta decay and neutrino mass,''
  Int.\ J.\ Mod.\ Phys.\ E {\bf 25} (2016) no.11,  1630007
  doi:10.1142/S0218301316300071
  [arXiv:1612.02924 [hep-ph]].
  %%CITATION = doi:10.1142/S0218301316300071;%%
%
%\cite{Guichon:2011gc}
\bibitem{Guichon:2011gc}
  P.~A.~M.~Guichon, A.~W.~Thomas and K.~Saito,
  %``Fermi matrix element with isospin breaking,''
  Phys.\ Lett.\ B {\bf 696} (2011) 536
  doi:10.1016/j.physletb.2011.01.005
  [arXiv:1101.2278 [nucl-th]].
%
%\cite{Ademollo:1964sr}
\bibitem{Ademollo:1964sr}
  M.~Ademollo and R.~Gatto,
  %``Nonrenormalization Theorem for the Strangeness Violating Vector Currents,''
  Phys.\ Rev.\ Lett.\  {\bf 13} (1964) 264.
  doi:10.1103/PhysRevLett.13.264
  %%CITATION = doi:10.1103/PhysRevLett.13.264;%%
%
%\cite{Garcia:1985xz}
\bibitem{Garcia:1985xz}
  A.~Garcia, P.~Kielanowski and A.~Bohm,
  %``The Beta Decay Of Hyperons,''
  Lect.\ Notes Phys.\  {\bf 222} (1985) 1.
  doi:10.1007/3-540-15184-2
  %%CITATION = doi:10.1007/3-540-15184-2;%%
%
\end{thebibliography}

\end{document}